\documentclass[aps, pra, twocolumn, superscriptaddress, floatfix, nofootinbib]{revtex4}

\usepackage{times, amsmath, amsthm, amssymb, graphicx, setspace, sistyle}
\pdfoutput=1

\def\citenum#1{{\def\@cite##1##2{##1}\cite{#1}}}

\newcommand{\ket}[1]{|#1\rangle}
\newcommand{\bra}[1]{\langle#1|}

\DeclareMathOperator{\tr}{Tr}

\newcommand{\arxiv}[1]{\href{http://arxiv.org/abs/#1}{\tt arXiv:\nolinkurl{#1}}}

\begin{document}

\title{Limits of quantum speedup in photosynthetic light harvesting}
\date{\today}
\author{Stephan Hoyer}
\email{shoyer@berkeley.edu}
\affiliation{Berkeley Quantum Information and Computation Center}
\affiliation{Department of Physics}
\author{Mohan Sarovar}
\author{K. Birgitta Whaley}
\email{whaley@berkeley.edu}
\affiliation{Berkeley Quantum Information and Computation Center}
\affiliation{Department of Chemistry\\University of California, Berkeley, CA, 94720, USA}

\begin{abstract}
It has been suggested that excitation transport in photosynthetic light harvesting complexes features speedups analogous to those found in quantum algorithms. Here we compare 
the dynamics in these light harvesting systems to 
the dynamics of quantum walks, in order to elucidate the limits of such quantum speedups. For the Fenna-Matthews-Olson (FMO) complex of green sulfur bacteria, we show that while there is indeed speedup at short times, this is short lived (70 fs) despite longer lived (ps) quantum coherence. Remarkably, this time scale is independent of the details of the decoherence model. More generally, we show that the distinguishing features of light-harvesting complexes 
not only limit the extent of quantum speedup but also reduce rates of diffusive transport.
These results suggest that quantum coherent effects in biological systems are optimized for efficiency or robustness rather than the more elusive goal of quantum speedup.
\end{abstract}


\maketitle

\section{Introduction}

In the initial stages of photosynthesis, energy collected from light is transferred across a network of chlorophyll molecules to reaction centers~\cite{Blankenship2002, Cheng2009}. Recent experimental evidence showing long lived quantum coherences in this energy transport in several photosynthetic light-harvesting complexes suggests that coherence may play an important role in the function of these systems~\cite{Engel2007, Hohjai2007, Ishizaki2009a, Calhoun2009, Collini2010, Panitchayangkoon2010}.
In particular, it has been hypothesized that excitation transport in such systems to feature speedups analogous to those found in quantum algorithms~\cite{Engel2007, Mohseni2008}.
These comments have attracted much interest 
from quantum information theorists~\cite{Mohseni2008, Rebentrost2008, Rebentrost2008a, Plenio2008, Caruso2009, Caruso2009a, Chin2009, Sarovar2009, Wilde2009, Bradler2009},
although clearly photosynthesis is not implementing unitary quantum search \cite{Mohseni2008}.
The most direct analogy to such transport is found in quantum walks, which form the basis of a powerful class of quantum algorithms including quantum search~\cite{Ambainis2003, Childs2003, Shenvi2003, Ambainis2007, Ambainis2007a, Childs2008a}.
Unlike idealized quantum walks, however, real light harvesting complexes are characterized by disorder, energy funnels and decoherence. Whether any quantum speedup can be found in this situation has remained unclear.

Quantum walks are an important tool for quantum algorithms ~\cite{Ambainis2003, Childs2003, Shenvi2003, Ambainis2007, Ambainis2007a, Childs2008a}.
On the line, quantum walks feature ballistic spreading, $\langle x^{2} \rangle \propto t^{2}$, compared to the diffusive spreading of a classical random walk, $\langle x^{2} \rangle \propto t$, where $\langle x^2 \rangle$ denotes the mean squared displacement.  
Moreover, because they use superpositions instead of classical mixtures of states, on any graph with enough symmetry to be mapped to a line, quantum walks spread along that line in linear time
-- even when classical spreading is exponentially slow~\cite{Childs2003}.  
We shall refer to this enhanced rate of spreading as a generic indicator of ``quantum speedup.''  Such quantum speedup is important in quantum information processing, where it can  lead to improved scaling of quantum algorithms relative to their classical alternatives. 
Examples of such algorithmic speedup include spatially structured search~\cite{Shenvi2003}, element 
distinctness~\cite{Ambainis2007} and evaluating {\sc and-or} formulas~\cite{Ambainis2007a}. Quantum walks even provide a universal implementation for quantum computing~\cite{Childs2008a}.

Quantum walks also constitute one of the simplest models for quantum transport on arbitrary graphs. As such they provide a theoretical framework for several physical processes, including the transfer of electronic excitations in photosynthetic light harvesting complexes~\cite{Engel2007, Mohseni2008}.
The closed system dynamics in a light-harvesting complex are generally well described by a tight-binding Hamiltonian that is restricted to the single excitation subspace~\cite{Cheng2009},
\begin{equation}
	H = \sum_n E_n \ket n \bra n +  \sum_{n\neq m} J_{nm} \ket{n}\bra{m}.
	\label{eq:tightbinding}
\end{equation}
Here $\ket n$ represents the state where the $n$th chromophore (site) is in its electronic excited state and all other chromophores are in their ground state. $E_{n}$ is the electronic transition energy of chromophore $n$ and $J_{mn}$ is the dipole-dipole coupling between chromophores $n$ and $m$. 
This is a more general variant of the Hamiltonian for standard continuous-time quantum walks, where $E_{n} = 0$ and $J_{mn} \in \{0,1\}$.
The salient differences of the general model, variable site energies and couplings, arise naturally from the structure and role of light-harvesting complexes.
Non-constant site energies can serve as energy funnels and can enable the complex to absorb at a broader range of frequencies, while variable  couplings between sites reflect their physical origin as dipole-dipole interactions. These differences yield excitation dynamics that can deviate significantly from quantum walks.

Here we study the key question of whether excitation transport on light-harvesting complexes shows quantum speedup.
Such quantum speedup would be necessary for any quantum algorithm that offers algorithmic speedup relative to a classical search of physical space. (Achieving true algorithmic speedup, i.e., improvement over the best classical algorithms, would also require suitable scaling of the space requirements~\cite{Blume-Kohout2002}.)
We address this with a study of the Fenna-Matthews-Olson (FMO) complex of green sulfur bacteria, a small and very well-characterized photosynthetic complex~\cite{Blankenship2002, Cheng2009, Fenna1975, Adolphs2006}, the same complex for which long lived quantum coherences were recently observed and suggested to reflect execution of a natural quantum search~\cite{Engel2007}.
We also consider the dynamics of transport along a theoretical model of an extended chain of chromophores to elucidate the systematic influence of variable site energies and dephasing on transport.

\section{Fenna-Matthews-Olson complex}

\begin{figure}
	\includegraphics[]{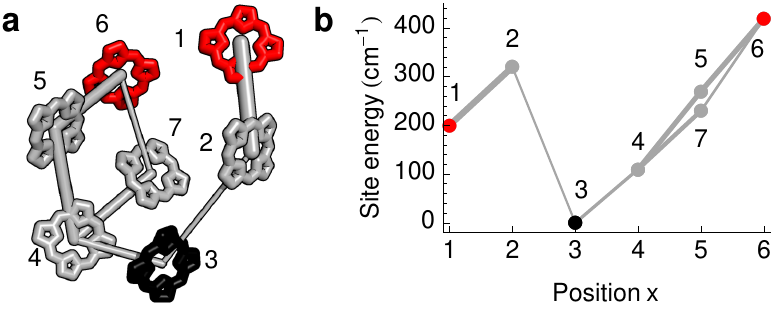}
	\caption{\label{fig:fmo-diagram}(a) Crystal structure of FMO complex of \emph{C.\ tepidum} (Protein Data Bank accession 3ENI), with lines between the chromophores representing dipolar couplings.  The thickness of the lines indicates the coupling strengths.  Only couplings above \SI{15}{cm^{-1}} are shown; the largest coupling is \SI{96}{cm^{-1}}. The full Hamiltonian is given in~\ref{sec:fmo-details}. 
	 (b) Site energies, shown relative to \SI{12210}{cm^{-1}}, in the reduced dimensionality model  derived from mapping the strongest couplings onto a one-dimensional graph. The red sites (1, 6) are source sites at which the excitation enters the complex and the black site (3) is the trap site from which the excitation is transferred to the reaction center~\cite{Adolphs2006, Wen2009}.}
\end{figure}

The FMO complex acts as a quantum wire, transporting excitations from a large, disordered antennae complex to a reaction center.
In addition to the possibility of quantum speedup~\cite{Engel2007,Mohseni2008}, recent studies have 
speculated that coherence may assist unidirectional transport along this wire~\cite{Ishizaki2009a} or suggested that it may contribute to overall efficiency~\cite{Rebentrost2008a}.
The crystal structure of FMO shows three identical subunits that are believed to function independently, each with seven bacteriochlorophyll-\textit{a} molecules embedded in a dynamic protein cage~\cite{Fenna1975}. 
A refined model Hamiltonian for the single excitation subspace is available from detailed quantum chemical calculations~\cite{Adolphs2006}, and the orientation of the complex was recently verified experimentally~\cite{Wen2009}.
By neglecting the weakest couplings in this model Hamiltonian, we see that transport in an individual monomer of FMO can be mapped to a one-dimensional path between chromophores, as shown in Figure~\ref{fig:fmo-diagram}.
This mapping of the excitation transport in FMO to a single dimension allows us to make contact with known results for quantum transport in one dimension and to quantitatively assess the extent of quantum speedup.

\section{Coherent dynamics}
\label{sec:coherent}

Under Hamiltonian dynamics, quantum walks on highly symmetric graphs can spread ballistically, but any significant lack of symmetry can lead to localization.  
Consider transport along the infinite line. Here, a random variation of any magnitude in the site energies leads to Anderson localization~\cite{Phillips1993}. Similarly, random variations in the coupling strengths contribute to localization~\cite{Fidder1991}. Systematic variations in site energies or coupling strengths can also cause localization~\cite{Serdyukova1992}, with the well-known instance of Bloch oscillations and Stark localization deriving from a linear bias in site energies~\cite{Hartmann2004}.
Both systematic and random variations in site energies remove the symmetry necessary for Bloch's theorem, so it is not surprising that their combination leads to localization as well~\cite{Luban1986, Kolovsky2008}.
Adding disorder into the graph structure also usually causes localization~\cite{Muelken2007} 
(although such disorder can also reduce localization if it makes an already disordered graph more connected~\cite{Giraud2005}).
The varied couplings, energy funnel and disordered energies evident in the FMO Hamiltonian 
depicted in Figure~\ref{fig:fmo-diagram} suggest that localization due to several of these effects will be significant for light-harvesting complexes.
The standard measure of
localization under coherent dynamics, the inverse participation ratio~\cite{Phillips1993} $\xi = \sum_i |\psi_i|^2/\sum_i |\psi_i|^4$, for the amplitudes $\psi_i$ in the site basis of an eigenstate $\psi$ of our model Hamiltonian confirms this intuition, since the typical eigenstate for FMO occupies only $\xi\sim 2$ sites (see also Ref.~\citenum{Brixner2005}).

We can estimate the timescale for localization $t_\text{loc}$ based upon the experimentally accessible parameters of geometry, localization length $\xi$ and an average coupling strength $J$. Consider the speed of the quantum walk as an upper bound on excitation transport speed prior to localization. For the continuous-time quantum walk on the infinite line, it is well known that $\langle x^{2} \rangle \sim 2 J^{2} t^{2}/\hbar^{2}$~\cite{Konno2005}. Similarly, starting at one end of an infinite line, $\langle x^{2} \rangle \sim 3 J^{2} t^{2}/\hbar^{2}$. This gives an average speed $g J/\hbar$, where $g=\sqrt 2$ and $g = \sqrt 3$ respectively, yielding the bound $t_{\text{loc}} \gtrsim \hbar\xi/g J$.
Our simulations of strict Hamiltonian dynamics (no decoherence) with FMO as described below find the onset of localization at $t_{\text{loc}} \sim \SI{70}{fs}$, which is close to the bound $\hbar\xi/g J \sim \SI{100}{fs}$ from the
 mean inverse participation ratio $\xi_\text{avg} \approx 2$, the mean coupling strength $J_\text{avg} \approx \SI{60}{cm^{-1}}$ and $g = \sqrt 3$ (this value for $g$ is suggested by the dominant pathways in Figure~\ref{fig:fmo-diagram} for transport starting at sites 1 or 6, which are the 
chromophores closest to the antenna complex~\cite{Adolphs2006, Wen2009}).

\section{Decoherence}

Decoherence, i.e.\ non-unitary quantum evolution, is also an essential feature of excitation dynamics in real systems.
In light harvesting complexes decoherence arises from interactions with the protein cage, the reaction center and the surrounding environment.
Recently it has been shown that some degree of fluctuation of electronic transition energies (i.e.\ dephasing in the site basis)
increases the efficiency of transport in FMO and other simple models otherwise limited by Anderson localization.
The intuition is that at low levels dephasing allows escape from localization by removing destructive interference, but at high levels it inhibits transport by inducing the quantum Zeno effect~\cite{Mohseni2008, Rebentrost2008, Plenio2008, Caruso2009}.

We consider here two essential types of decoherence: dephasing in the site basis and loss of excitations.
 Under the Born--Markov approximation, time evolution with this decoherence model is determined by the system Hamiltonian $H$ and a sum of Lindblad operators $\mathcal{L}$ according to~\cite{Breuer2002}
\begin{equation}
	\frac{\partial\rho}{\partial t} = -\frac{i}{\hbar} [H, \rho] + \mathcal{L}_{\text{loss}}(\rho) + \mathcal{L}_{\text{deph}}(\rho).
	\label{eq:master}
\end{equation}
$\mathcal{L}_{\text{loss}}$ describes loss of excitations with site dependent rates $\gamma_{n}$, including trapping to a reaction center, and is specified by the operator
\begin{equation}
	\bra n \mathcal{L}_{\text{loss}}(\rho) \ket m = - \frac{\gamma_{n} + \gamma_{m}}{2} \bra n \rho \ket m,
\end{equation}
which results in exponential decay of the diagonal elements of the density matrix $\rho$ (and corresponding decay of off-diagonal elements, which ensures positivity).
$\mathcal{L}_{\text{deph}}$ describes dephasing, which is specified by the operator
\begin{equation}
	\bra n \mathcal{L}_{\text{deph}}(\rho) \ket m = - \frac{\Gamma_{n} + \Gamma_{m}}{2} (1 - \delta_{nm}) \bra n \rho \ket m
\end{equation}
with site dependent dephasing rates $\Gamma_{n}$, and yields exponential decay of the off-diagonal elements of $\rho$. 

A remarkable feature of our results for FMO is that, as we shall demonstrate, they are independent of the finer details of the bath dynamics and system-bath coupling.
Thus for simplicity, we specialize here to the Haken-Strobl model~\cite{Haken1967} and restrict losses to trapping by the reaction center. 
The ease of calculations with the Haken-Strobl model has made it popular for simulating the dynamics of light harvesting complexes~\cite{Rebentrost2008, Plenio2008, Caruso2009, Leegwater1996, Gaab2004}.
Following Rebentrost et al.~\cite{Rebentrost2008}, we use the spatially uniform, temperature dependent dephasing rate $\Gamma = 2 \pi k_{B} T E_{R} /\hbar^{2} \omega_{c}$, for an ohmic spectral density with bath reorganization energy $E_{R} = \SI{35}{cm^{-1}}$ and cut-off frequency $\omega_{c} = \SI{150}{cm^{-1}}$.
This dephasing rate holds when chromophores are treated as qubits coupled to independent reservoirs in the spin-boson model, in the thermal (or Markovian) regime $t \gg \hbar/k_B T$~\cite{Breuer2002}, which corresponds to $t \gg \SI{25}{fs}$ at \SI{300}{K}. Although 
the exact parameters of the protein environment surrounding
FMO are unclear~\cite{Ishizaki2009a}, these are reasonable estimates.
This model gives dephasing rates
$(\SI{69}{fs})^{-1}$ at \SI{77}{K} and $(\SI{18}{fs})^{-1}$ at \SI{300}{K}.
Also, as in Ref.~\citenum{Rebentrost2008} we restrict trapping to site 3 at a rate of $\gamma_{3}=\SI{1}{ps^{-1}}$. Finally, we neglect exciton recombination since it occurs on much slower timescales ($\sim$\SI{1}{ns^{-1}}) and in any case it does not alter the mean-squared displacement. Therefore, we set $\gamma_{i}=0$ for $i\neq 3$.
While more realistic decoherence models would incorporate thermal relaxation, spatial and temporal correlations in the bath and strong system-bath coupling~\cite{Ishizaki2009}, this treatment is sufficient for analyzing 
the quantum speedup, as justified in detail by comparison with the corresponding analysis using more realistic simulations of FMO~\cite{Ishizaki2009} in~\ref{sec:aki-model}.

\section{Limits of quantum speedup}

To access the extent of quantum speedup for a system that can be mapped to a line, a natural measure is the exponent $b$ of the power law for the mean squared displacement $\langle x^{2} \rangle \propto t^{b}$.
In particular, we use the best-fit exponent $b$ from the slope of the log-log plot of the mean squared displacement $\langle x^2 \rangle = \tr[\rho x^2]/ \tr \rho$ versus $t$. The value $b=1$ corresponds to the limit of diffusive transport, whereas $b=2$ corresponds to ideal quantum speedup as in a quantum walk (ballistic transport).

\begin{figure}
	\includegraphics[]{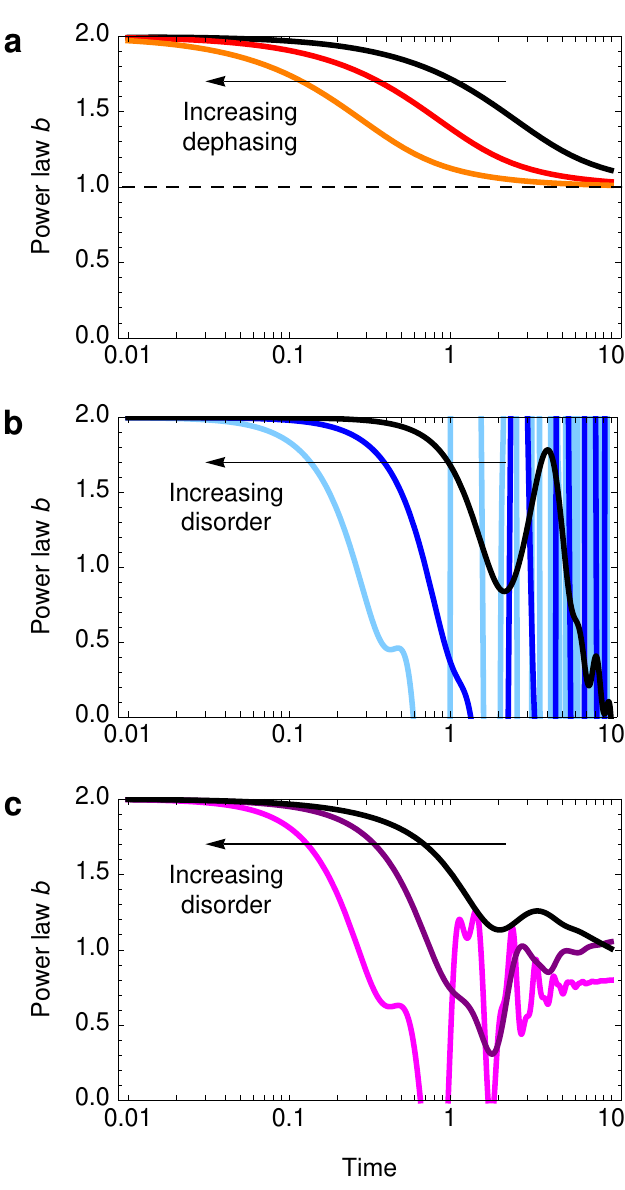}
	\caption{\label{fig:powerlaws}The best fit power law exponent $b$ for the mean-squared displacement with $J/\hbar=1$, for transport along a linear chain with either (a) increasing dephasing ($\Gamma = 1,3,9$) and no static disorder ($E_n = 0$), (b) no dephasing ($\Gamma=0$) and increasing static disorder (independent site energies $E_n$ from a single Gaussian distribution normalized to standard deviation $\sigma = 1,3,9$)
	or (c) same disorder as (b), 
	but with finite dephasing rate $\Gamma = 1$. Panels (b) and (c) show results for single instances of disorder. When an ensemble average over different realizations of static disorder is taken, the power law varies smoothly instead of oscillating as in (b) and (c).}
\end{figure}

The timescale for quantum speedup is generally bounded above by both the timescale for dephasing, $t_{\textrm{deph}} = 1/\Gamma$, and the timescale for static disorder to cause localization, $t_{\textrm{loc}}$. To see this, it is illustrative to consider transport along a linear chain with constant nearest neighbor couplings $J$.
For dephasing but no static disorder 
or energy gradient  ($E_n = 0$), the mean squared displacement shows a smooth transition from ballistic to diffusive transport~\cite{Schwarzer1972, Grover1971} \footnote{This result, obtained in the context of excitation transport in molecular crystals, also applies to decoherent continuous-time quantum walks since they use the same tight-binding Hamiltonian.},
\begin{equation}
	\langle x^2 \rangle  = \frac{4 J^2}{\hbar^2 \Gamma} \left[t + \frac{1}{\Gamma} \left(1 - e^{- \Gamma t}\right)\right].
	\label{eq:displacementalltimes}
\end{equation}
The corresponding power law transition is shown in Figure~\ref{fig:powerlaws}a.
Technically, transport remains super-diffusive even after $t_\text{deph}$, since the power never drops below $b=1$. In contrast, for static disorder but no dephasing, there is a sudden transition from ballistic transport to essentially no transport at all (localization). The power law starts at $b=2$, then drops and begins to oscillate wildly after $t_\text{loc}$ as the wave function continues to evolve in a confined region. Several examples are shown in Figure~\ref{fig:powerlaws}b for variable strengths of disorder. Figure ~\ref{fig:powerlaws}c shows the behavior with both strong static disorder and dephasing, as in light harvesting complexes. 
In this case, the transport can even exhibit a sub-diffusive power law. The reason for this sub-diffusive behavior may be easiest to understand by analogy to the Anderson model
 (random site energies) in an infinite chain. Consider transport in such a system under weak dephasing and with an ensemble average over different realizations of strong static disorder. In this case, one expects transport must transition from ballistic ($b=2$) at short times, to localized ($b=0$) at intermediate times, to diffusive at long times ($b=1$). (The situation at long times is analyzed more explicitly in Section~\ref{sec:diffusive} and~\ref{sec:decaycoherence}.)
  For stronger dephasing as in Figure~\ref{fig:powerlaws}c (and as we shall 
  see with FMO), the fully localized regime with no transport may never be realized, but transport will still be sub-diffusive for intermediate times.

\begin{figure*}
	\includegraphics{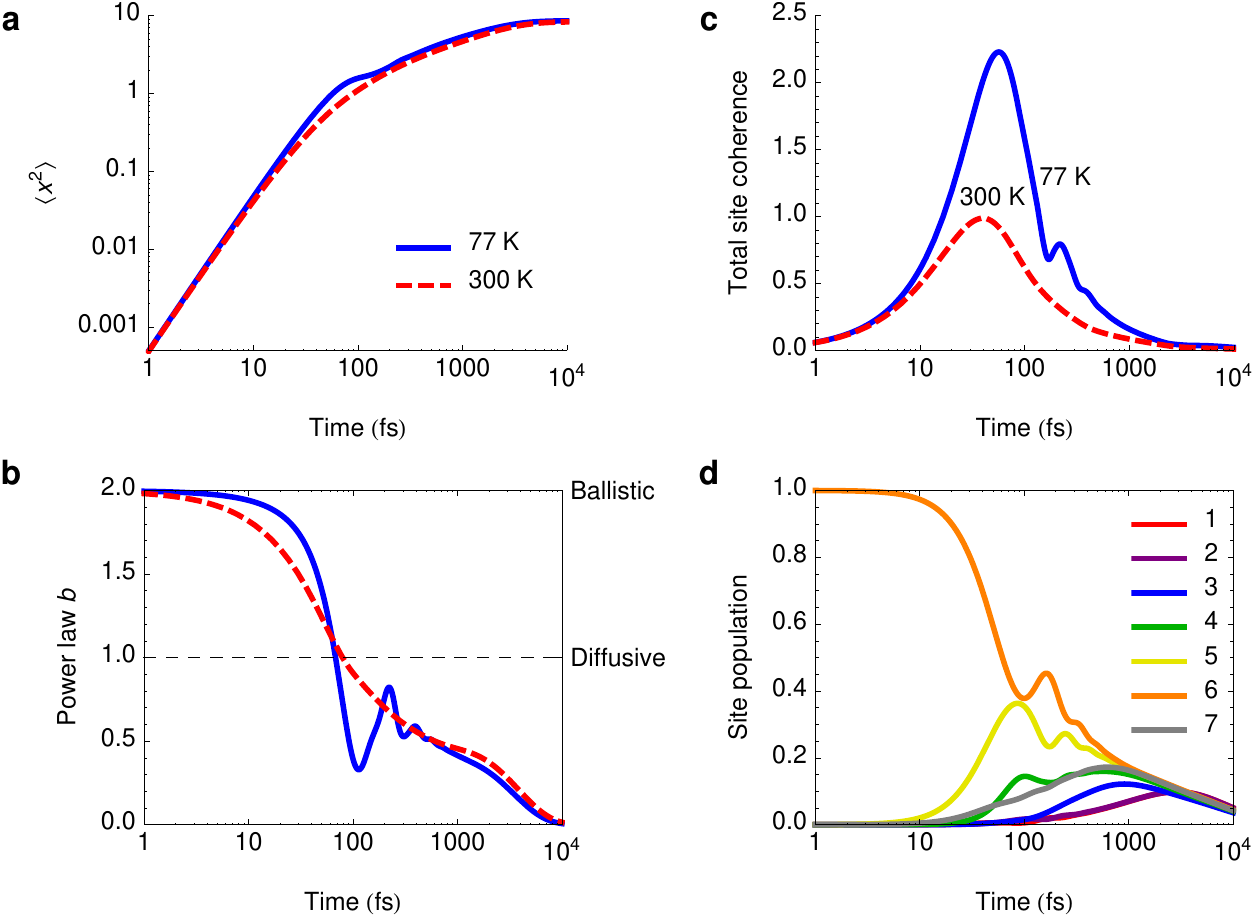}
	\caption{\label{fig:fmo-transport}Results of simulations on FMO with the decoherence model described in the text. (a) Log-log plot of mean-squared displacement $\langle x^{2} \rangle$ as a function of time, for an initial excitation at site 6. (b) Power law exponent for the mean-squared displacement, given by the slope of the plot in panel a. The dashed line $b=1$ separates super- and sub-diffusive transport (see text). (c) Total site coherence, given by the sum of the absolute value of all off-diagonal elements of the density matrix in the site basis, indicates persistent coherence for \SI{\sim 500}{fs}. (d) Oscillating site populations (shown for \SI{77}{K} only) also indicate persistent coherence for hundreds of femtoseconds.}
\end{figure*}

Figure~\ref{fig:fmo-transport} shows the results of a simulation of FMO dynamics 
made with our simple decoherence model. The displacement $x_i$ of a site $i$ is given by its position in the one-dimensional mapping 
that is presented in Figure~\ref{fig:fmo-diagram}b. Note that our simulations use the full Hamiltonian; the one-dimensional map is only used in the analysis of the results, to determine the displacement $x_i$ of each site.
The upper-left panel, (a), shows the 
mean squared displacement $\langle x^2 \rangle$ of an excitation initially localized on site 6, 
one of the two excitation source sites, and the lower-left panel, (b), the best fit exponent $b$ for the power law $\langle x^{2} \rangle \propto t^{b}$. 
Our results show that even though coherence in this simple model lasts for \SI{\sim 500}{fs} (see Figure~\ref{fig:fmo-transport}c), a transition from initially ballistic to sub-diffusive transport occurs after only \SI{\sim 70}{fs}, independent of the dephasing rate.
While increased dephasing causes a faster initial decrease of the power law, dephasing alone
cannot lead to sub-diffusive transport, as is clear from Equation \eqref{eq:displacementalltimes}.
We note that the transition to sub-diffusive transport at \SI{70}{fs} occurs at the same time-scale as the onset of localization
(Section~\ref{sec:coherent}), implying that even though there is 
persistent coherence beyond this time, it no longer yields quantum speedup because of static disorder.
The sub-diffusive power law at intermediate times 
(\SI{\sim 100}{fs} -- \SI{2}{ps}) arises from the interplay of this disorder-induced localization and dephasing 
that is discussed above for the linear chain model.  At longer times the power law exponent $b$ goes to zero because of the finite size of the system.
Since complete energy transfer through FMO takes picoseconds, 
this analysis shows that most of the excitation transport is 
formally sub-diffusive.

While FMO is a relatively small system, so that terms such as ``ballistic'' and ``diffusive'' cannot literally describe transport across its seven chromophores, this time dependence of the power law of spreading would also characterize larger artificial or natural systems.
Our results here are robust to variations in the strength of the trap, and to whether the initial excitation is at site 1 or 6, the sites believed to be the primary source for excitations in FMO~\cite{Adolphs2006}.
They are also independent to variations of the map to a one-dimensional system in Figure~\ref{fig:fmo-diagram}, such as counting the number of chromophores away from the trap site instead of along the entire line, or using the real space distance between adjacent chromophores instead of assuming that each pair is 
merely separated by a unit lattice distance.

Our timescale for short-lived quantum speedup in FMO should be largely independent of the details of the decoherence model, since the non-uniform energy landscape will always limit ballistic transport to times prior to $t_\text{loc}\sim \SI{70}{fs}$. This is confirmed by applying our analysis to the results of recent calculations for FMO made with a considerably more sophisticated decoherence model that incorporates thermal relaxation, temporal correlations in the bath and strong system-bath coupling~\cite{Ishizaki2009}.  As demonstrated in~\ref{sec:aki-model},
these more realistic calculations also predict a loss of quantum speedup after \SI{70}{fs}.

\section{Diffusive transport}
\label{sec:diffusive}

\begin{figure}
	\includegraphics[]{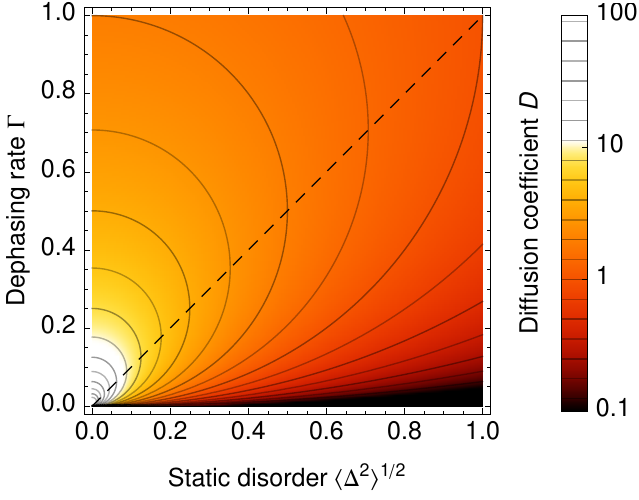}
	\caption{\label{fig:diffusion-coeffs}Diffusion coefficients at long times as a function of a constant dephasing rate $\Gamma$ and the static disorder between adjacent sites, as calculated by \eqref{eq:diffusion-simplified} with $J=\hbar=1$. Solid lines are contours; the dashed line is the optimal dephasing rate for given static disorder. The ideal quantum walk, which is non-diffusive, is at the origin.}
\end{figure}

To gain general insight into the interplay between disorder and dephasing
for light harvesting complexes, we now consider the dynamics of an infinite linear chain at long times with variable site energies, couplings and dephasing rates.
By extending an analysis for damped Bloch oscillations~\cite{Kolovsky2002}, we find that any set of non-zero dephasing rates $\Gamma_{n}$ asymptotically leads to diffusive transport $\langle x^{2} \rangle \sim 2 D t$.
Haken and Reineker also performed a reduction of the Haken-Strobl model without disorder to a diffusion equation along similar lines~\cite{Haken1972}. 
Note that since the Haken-Strobl model does not include energy relaxation, no classical drift velocity will be obtained from these dynamics.
The time evolution of an arbitrary density matrix element for an infinite linear chain under site dependent dephasing rates is given by
\begin{multline}
	\dot \rho_{nm} = -\frac{i}{\hbar} (J_{n-1} \rho_{n-1,m} + J_{n} \rho_{n+1,m} \\
	 - J_{m} \rho_{n,m+1} - J_{m-1} \rho_{n,m-1}) \\
	- \left[ \frac{i}{\hbar} (E_{n} - E_{m}) + \frac{\Gamma_{n} + \Gamma_{m}}{2} (1 - \delta_{nm}) \right] \rho_{n,m},
	\label{eq:rhonm}
\end{multline}
where $J_{n} \equiv J_{n,n+1}$.
If we neglect terms in the second off-diagonal relative to the diagonal terms (see justification in~\ref{sec:decaycoherence}) we obtain
\begin{align}
	\dot \rho_{n+1,n} &= - \frac{i}{\hbar} J_{n}(\rho_{n,n} - \rho_{n+1,n+1}) - ( \frac{i}{\hbar} \Delta_{n} + \Gamma^{\prime}_{n}) \rho_{n+1,n},
\end{align}
where $\Delta_{n} \equiv E_{n+1} - E_{n}$ and $\Gamma^{\prime}_{n} \equiv (\Gamma_{n} + \Gamma_{n+1})/2$.
This equation has the solution
\begin{align}
	\rho_{n+1,n} =&  \frac{i}{\hbar} e^{-(i \Delta_{n}/\hbar + \Gamma_{n}) t} \notag \\
	&\times \int_{0}^{t} J_{n} (\rho_{n+1, n+1} - \rho_{n,n}) e^{(i \Delta_{n}/\hbar + \Gamma_{n}) t^{\prime}} dt^{\prime} \\
	\approx& \frac{J_{n} (\rho_{n+1,n+1} - \rho_{n,n})}{\Delta_{n} - i \hbar \Gamma^\prime_{n}},
	\label{eq:offdiag}
\end{align}
in the nearly stationary regime where site populations vary slowly compared to $1/\Gamma^{\prime}_{n}$ and $\hbar/\Delta_{n}$.
Inserting \eqref{eq:offdiag} and the corresponding result for $\rho_{n-1,n}$ into \eqref{eq:rhonm} we obtain
\begin{multline}
	\dot \rho_{nn} = \frac{2 J_{n}^{2} \Gamma^{\prime}_{n}}{\Delta_{n}^{2} + \hbar^{2}{\Gamma^{\prime}}^{2}_{n}} (\rho_{n+1,n+1} - \rho_{nn}) \\
	+ \frac{2 J_{n-1}^{2} {\Gamma^{\prime}}_{n-1}}{\Delta_{n-1}^{2} + \hbar^{2} {\Gamma^{\prime}}_{n-1}^{2}} (\rho_{n-1,n-1} - \rho_{nn}).
	\label{eq:fullevolve}
\end{multline}
This is now a classical random walk with variable bond strengths between sites.
For asymptotically large times it has been 
proven to approximate the diffusion equation with 
coefficient given by
\begin{equation}
	D = \left\langle \frac{\Delta_{n}^{2} + \hbar^{2} {\Gamma^{\prime}_{n}}^{2}}{2 J_{n}^{2} \Gamma^{\prime}_{n}} \right\rangle^{-1},
	\label{eq:diffusion-coeff}
\end{equation}
as long as the average over sites in the diffusion coefficient is well-defined~\cite{Hughes1996}.
With $\Delta_{n}$, $J_{n}$ and $\Gamma_{n}$ constant, this matches the known diffusion coefficient 
\cite{Kolovsky2002, Dunlap1988}.
For constant coupling $J_{n} = J$ and constant dephasing $\Gamma^{\prime}_{n} = \Gamma$, we note that \eqref{eq:diffusion-coeff} reduces to
\begin{equation}
	D = \frac{2 J^{2} \Gamma}{\langle \Delta_{n}^{2} \rangle + \hbar^{2} \Gamma^{2}},
	\label{eq:diffusion-simplified}
\end{equation}
which is plotted
as a function of $\langle \Delta_{n}^{2} \rangle^{1/2}$ and $\Gamma$ in Figure~\ref{fig:diffusion-coeffs} \footnote{We note that the parameter $\langle \Delta_{n}^{2} \rangle^{1/2}$ does not completely specify the degree of localization, although it does show that Stark and Anderson localization equivalently influence diffusive transport. For example, the inverse participation ratio in these two cases differs, even with the same value of $\langle \Delta_{n}^{2} \rangle^{1/2}$.}.
Note that with $\langle \Delta_{n}^{2} \rangle=0$, this is the long time limit of \eqref{eq:displacementalltimes}.
In the case of Equation \eqref{eq:diffusion-simplified}, for a given degree of static disorder, we find an optimal dephasing rate $\hbar\Gamma = \langle \Delta_{n}^{2} \rangle^{1/2}$.
However, Figure~\ref{fig:diffusion-coeffs} shows that for a given dephasing rate, 
excitation transport would be improved by reducing the static disorder 
to further delocalize the system.

\section{Conclusions}

These principles show that quantum speedup across a photosynthetic or engineered system requires not
only long-lived quantum coherences but also excitons delocalized over the entire complex. Such completely delocalized excitons do not exist in FMO and are also unlikely in other light harvesting complexes because of the presence of energy gradients and disorder.
Moreover, Equation \eqref{eq:diffusion-simplified} makes it clear that the conditions needed for longer lived quantum speedup (reduced dephasing and static disorder) are those necessary for faster diffusive transport as well.

The short-lived nature of quantum speedup in light harvesting complexes that we have established here implies that the natural process of energy transfer across these complexes does not correspond to a quantum search.
Indeed, neither a 
formal quantum nor classical search may be necessary, since pre-determined (evolved) energy gradients can guide 
relaxation to reaction centers, even though such gradients suppress coherent and dephasing-assisted transport.
This is particularly relevant for systems such as FMO which either receive excitations one at a time or have isolated reaction centers.
Instead of yielding dynamical speedup like that in quantum walk algorithms, quantum coherence in photosynthetic light harvesting appears more likely to contribute to other aspects of transport, such as overall efficiency or robustness.  
We emphasize that a restricted extent of quantum speedup does not 
 imply that there is no significant quantum advantage due to long-lived coherence in electronic excitation energy transfer.
Identifying the specific nature of any ``quantum advantage'' for FMO will clearly require more detailed analysis of the dynamics, particularly in the sub-diffusive regime. 
Related examples of such ``quantum advantage" are found in the LH2 complex of purple bacteria, where coherence has been specifically shown to improve 
both the speed~\cite{Jang2004} and robustness~\cite{Cheng2006} of transport from the B800 to B850 ring.

\section*{Acknowledgments}

We thank Yuan-Chung Cheng, Graham R.\ Fleming, Akihito Ishizaki, Drew Ringsmuth and Robert J.\ Silbey for discussions, and Ahikito Ishizaki for
providing us with the results of his calculations on FMO from Ref.~\citenum{Sarovar2009}. This work was supported by DARPA under Award No.~N66001-09-1-2026.

\subsection*{Note added}

While revising this manuscript, a reduction of the Haken-Strobl model to a classical random walk similar to the one we perform in Section \ref{sec:diffusive} was published by Cao and Silbey~\cite{Cao2009}.

\appendix

\section{FMO Hamiltonian}
\label{sec:fmo-details}

In all of our calculations, we use the Hamiltonian calculated for \emph{C.\ tepidum} by Adolphs and Renger~\cite{Adolphs2006}. In the site basis, this is given by
\begin{equation}
	\label{eq:HFMO}
	\left(\begin{array}{ccccccc}{\bf 200} & {\bf -96} & 5 & -4.4 & 4.7 & -12.6 & -6.2 \\
	{\bf -96} & {\bf 320} & {\bf 33.1} & 6.8 & 4.5 & 7.4 & -0.3 \\
	5 & {\bf 33.1} & {\bf 0} & {\bf -51.1} & 0.8 & -8.4 & 7.6 \\
	-4.4 & 6.8 & {\bf -51.1} & {\bf 110} & {\bf -76.6} & -14.2 & {\bf -67} \\
	4.7 & 4.5 & 0.8 & {\bf -76.6} & {\bf 270} & {\bf 78.3} & -0.1 \\
	-12.6 & 7.4 & -8.4 & -14.2 & {\bf 78.3} & {\bf 420} & {\bf 38.3} \\
	-6.2 & -0.3 & 7.6 & {\bf -67} & -0.1 & {\bf 38.3} & {\bf 230} \end{array}\right),
\end{equation}
with units of cm$^{-1}$ and a total offset of \SI{12210}{cm^{-1}}.
Bold entries indicate those shown in Figure~\ref{fig:fmo-diagram}.
In units with $\hbar=1$, we note that the rate $\SI{1}{ps^{-1}} \equiv  \SI{5.3}{cm^{-1}}$.

\section{Reduced hierarchy equations model for FMO}
\label{sec:aki-model}
Energy transfer dynamics in photosynthetic complexes can be difficult to model because perturbations from the surrounding protein environment can be large, and the timescale of the protein dynamics 
is similar to the timescales of excitation transport. This makes common approximations involving perturbative treatments of system-bath coupling and Markovian assumptions on the bath invalid. Recently a non-perturbative, non-Markovian treatment of energy transfer has been formulated by Ishizaki and Fleming~\cite{Ishizaki2009}.
This model assumes: (i) a bilinear exciton-phonon coupling, (ii) protein fluctuations that are described by Gaussian processes, (iii) a factorizable initial state of chromophores and protein environment, (iv) protein fluctuations that are exponentially correlated in time, and (v) no spatial correlations of fluctuations.

To verify our results in Figure~\ref{fig:fmo-transport} of the main paper with this more realistic model for excitation dynamics, we 
apply the power law analysis of mean squared displacement to the results of the 
simulations that were used to calculate entanglement dynamics in Ref.~\citenum{Sarovar2009} using the non-perturbative non-Markovian method of Ref.~\citenum{Ishizaki2009a}.   
\begin{figure}[]
	\includegraphics[]{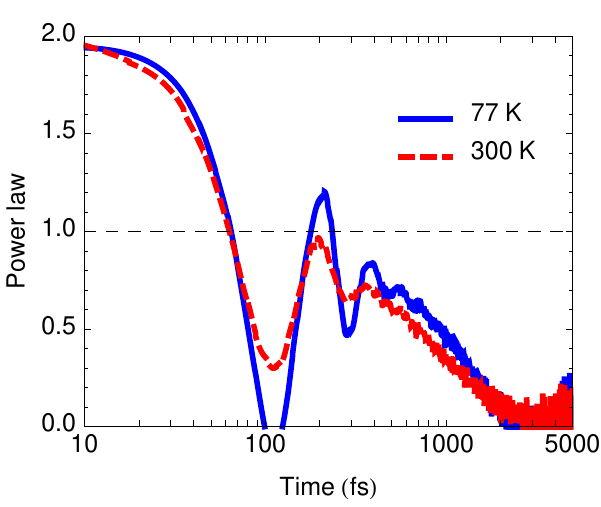}
	\caption{\label{fig:aki-figure}For the reduced hierarchy equation model as applied to FMO, we plot the power law for the mean squared displacement as in Figure~\ref{fig:fmo-transport}b. The plot is interpolated from a fourth order polynomial fit of density matrix elements for results from a numerical simulation on FMO with sampling approximately every \SI{8}{fs}, with the initial excitation at site 6
	\cite{Ishizaki2009a}.}
\end{figure}
These simulations used a reorganization energy of the protein environment of \SI{35}{cm^{-1}}, phonon relaxation time of \SI{100}{fs}, and a reaction center trapping rate of $(\SI{4}{ps})^{-1}$, all of which are consistent with literature on FMO~\cite{Adolphs2006, Cho2005}.  Figure~\ref{fig:aki-figure} shows the power law analysis for 2 different temperatures, \SI{77}{K} and \SI{300}{K}.
Our analysis of these realistic simulations show that longer lasting coherences (as discussed in Refs.~\citenum{Ishizaki2009a, Ishizaki2009}) are evident in the power law oscillations, but that transport is still nevertheless sub-diffusive after \SI{\sim 70}{fs},
for both temperatures.

It is particularly noteworthy that our results for FMO hold even under a model incorporating finite temperature relaxation.
In principle, thermal relaxation could bias the dynamics on a linear chain with a classical drift velocity~\cite{Ponomarev2006} to give the power law $\langle x^2 \rangle \propto t^2$ even without quantum speedup. However, the above 
analysis of FMO simulations made with the most realistic treatment of relaxation available today (Figure~\ref{fig:aki-figure}) shows that this is not the case in this system.

\section{Decay of coherences in linear transport}
\label{sec:decaycoherence}
For a constant dephasing rate $\Gamma \gg J/\hbar$, there is a simple analytical argument that the off-diagonal elements of the density matrix in the site basis can be neglected relative to the main diagonal~\cite{Fedichkin2006}. 
In the case of transport along a line, application of the analysis of Ref.~\citenum{Fedichkin2006} to Equation \eqref{eq:rhonm} shows that the first off-diagonal elements decay at rate $\Gamma$, unless more coherence is generated from the main diagonal. 
If the main diagonal elements have order 1, then the first off-diagonal elements cannot have magnitude greater than order $J/\hbar\Gamma$. A similar argument bounds the second diagonal elements as less than $(J/\hbar\Gamma)^{2}$ and so forth. For large $\Gamma$, this justifies neglecting this second off-diagonal relative to the main diagonal. The higher order terms in this expansion can also be calculated explicitly \cite{Cao2009}.

\begin{figure*}[]
	\includegraphics[]{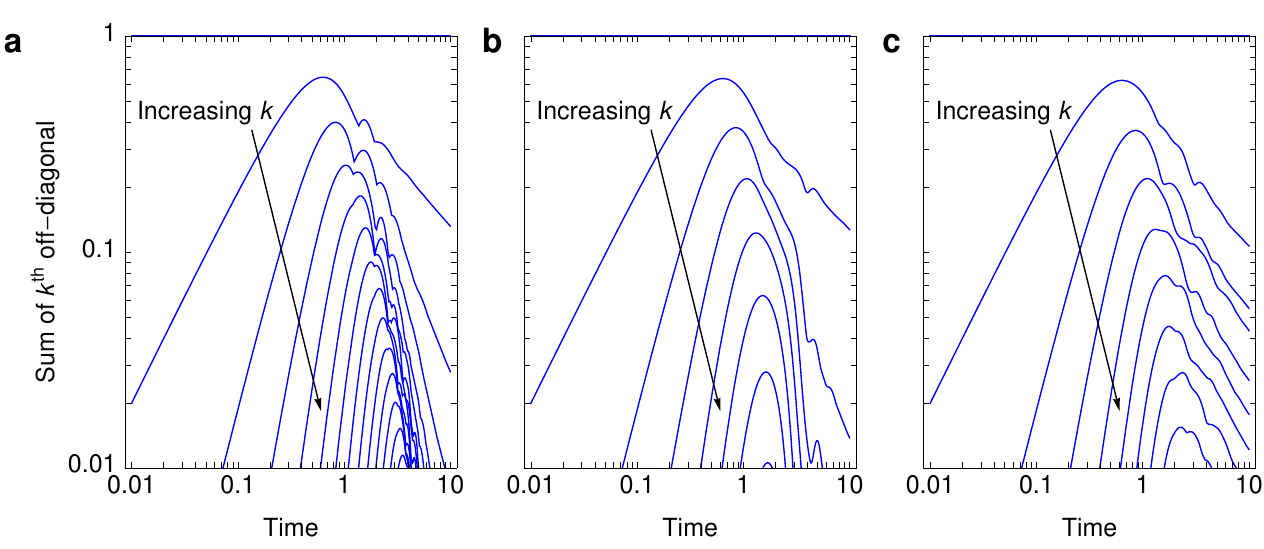}
	\caption{\label{fig:coherence-decay}\textbf{Decay of coherences.} Log-log plot of the sum of the absolute values $k$th off-diagonal elements of the density matrix over time for constant dephasing and coupling strengths $\Gamma=J/\hbar=1$. Panels show the quantum walk with dephasing (\textbf{a}), and Stark (\textbf{b}) and Anderson (\textbf{c}) localization with dephasing and $\langle \Delta_{n}^{2} \rangle^{1/2} = \pi/2$. Note that the main diagonal always sums to 1.}
\end{figure*}

From extensive numerical experiments with infinite chains, we have found uniformly that the off-diagonal elements of the density matrix decay nearly monotonically after time $1/\Gamma$ even when $\Gamma \gg J/\hbar$ does not hold, although an analytic proof of this result has eluded us. Several examples are presented in Figure~\ref{fig:coherence-decay}. This matches the analytical result for finite systems in the Haken-Strobl model that coherence must vanish eventually for any non-zero $\Gamma$~\cite{Parris1988}.


\bibliographystyle{../h-physrev-titles}
\bibliography{../quantumphoto}

\end{document}